\documentclass[12pt]{article}
\usepackage{graphicx}
\begin{document}
\bigskip

\centerline{\bf Ferromagnetic Phase Transition in Barab\'asi-Albert Networks}

\bigskip

Agata Aleksiejuk*, Janusz A. Ho{\l}yst* and Dietrich Stauffer

\bigskip
\noindent
Institute for Theoretical Physics, Cologne University, D-50923 K\"oln, Euroland

\bigskip
\noindent
* Permanently at: Faculty of Physics, Warsaw University of Technology, 
Koszykowa 75, PL-00--662 Warsaw, Poland

\bigskip

\noindent
Abstract:
Ising spins put onto a Barab\'asi-Albert scale-free network show an effective
phase transition from ferromagnetism to paramagnetism upon heating, with an
effective critical temperature increasing as the logarithm of the system size. 
Starting with
all spins up and upon equilibration pinning the few most-connected spins down 
nucleates the phase with most of the spins down.


\bigskip
\bigskip

Networks with more complicated connectivities than periodic lattices have been
investigated in detail recently. For example, the Barab\'asi-Albert network
\cite{barabasi} is grown such that the probability of a new site to be connected
to one of the already existing sites is proportional to the number of the 
previous connections to this already existing site: The rich get richer. Similar
networks have been investigated \cite{internet} to check if destruction of 
a few computers will split the percolating cluster of the internet (i.e. the
set of all computers in the world connected directly or indirectly with each
other). Networks exist also in social models where vertices are individuals 
or organizations, and links correspond to social relationships between them
\cite{wassermann}. However, as far as we know, all studies of the scale-free
Barab\'asi-Albert networks considered only the {\it topology} and no
{\it interactions} between linked vertices.

Here we investigate the ordering
phenomenon in this Barab\'asi-Albert network, if Ising spins are put onto the
sites (vertices) of the network. 
We assume ferromagnetic coupling between linked spins and positive temperature
$T$ of the system.  Such a magnet
would show paramagnetism if the whole network is still connected but only weakly
such that thermal fluctuations destroy the spontaneous magnetization. 
(Bose-Einstein condensation on similar networks was already studied before 
\cite{bose}, as was the Ising model on small-world networks \cite{pekalski}, and
other Ising models are in preparation \cite{vicsek}.)
In the social example one could identify exp($-{\rm const}/T$) with the 
probability that members of the same social group are not convinced to share 
the same opinion. (Spin variables $S_i = \pm 1$ then correspond to two possible 
opinions of the group members on the same subject.)

\begin{figure}[hbt]
\begin{center}
\includegraphics[angle=-90,scale=0.45]{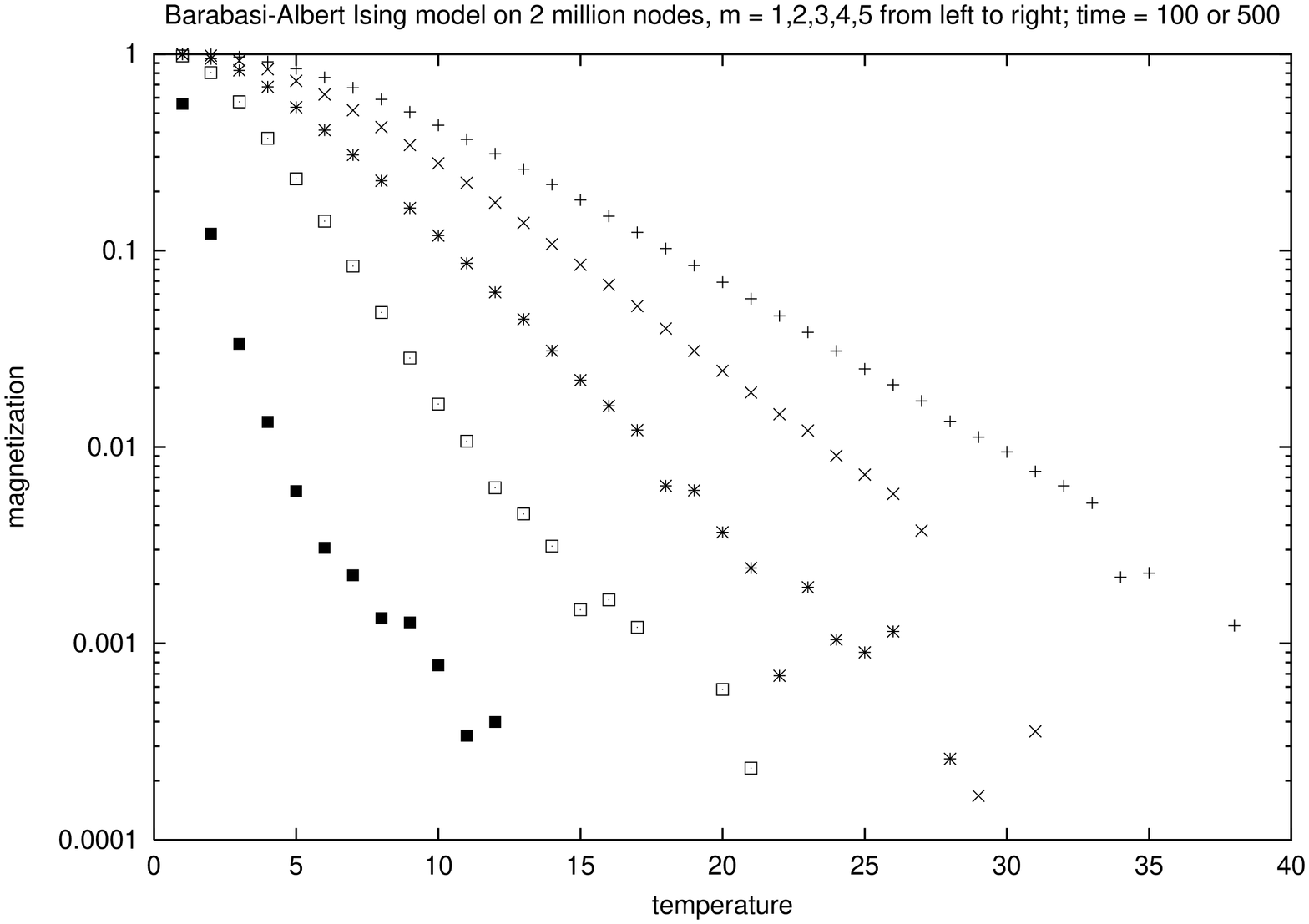}
\includegraphics[angle=-90,scale=0.45]{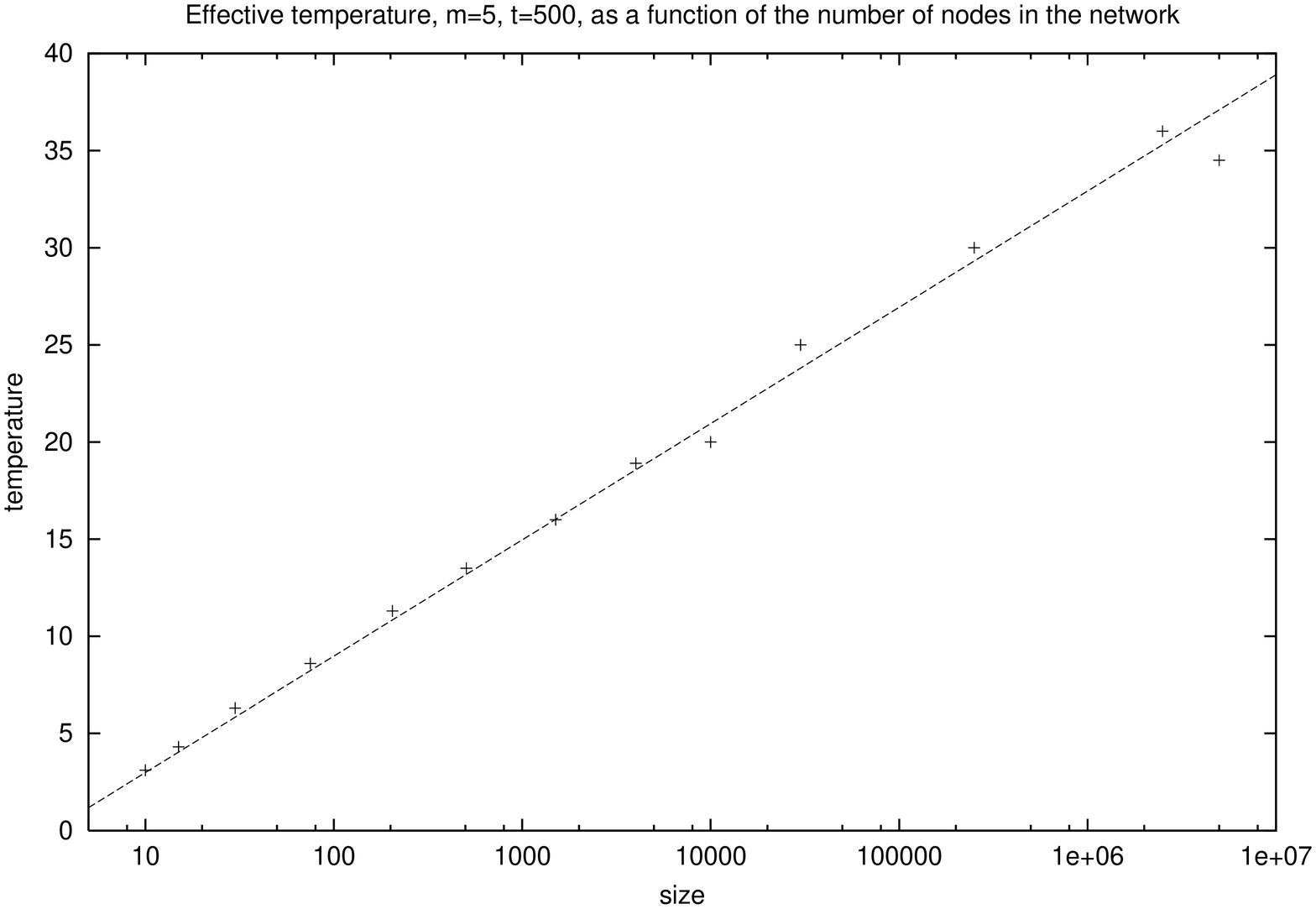}
\end{center}
\caption{
(a) $M$ versus temperature for 2 million nodes and various $m$, for $250 \le t 
\le 500$ (shorter times $t$ were used far below $T_c$).
(For $m=1$ even 60 million nodes were simulated.)
(b) Effective $T_c$ versus $m+N$ for $m=5$ and various $N$, averaged over up 
to 1000 samples.
}
\end{figure}  

\begin{figure}[hbt]
\begin{center}
\includegraphics[angle=-90,scale=0.5]{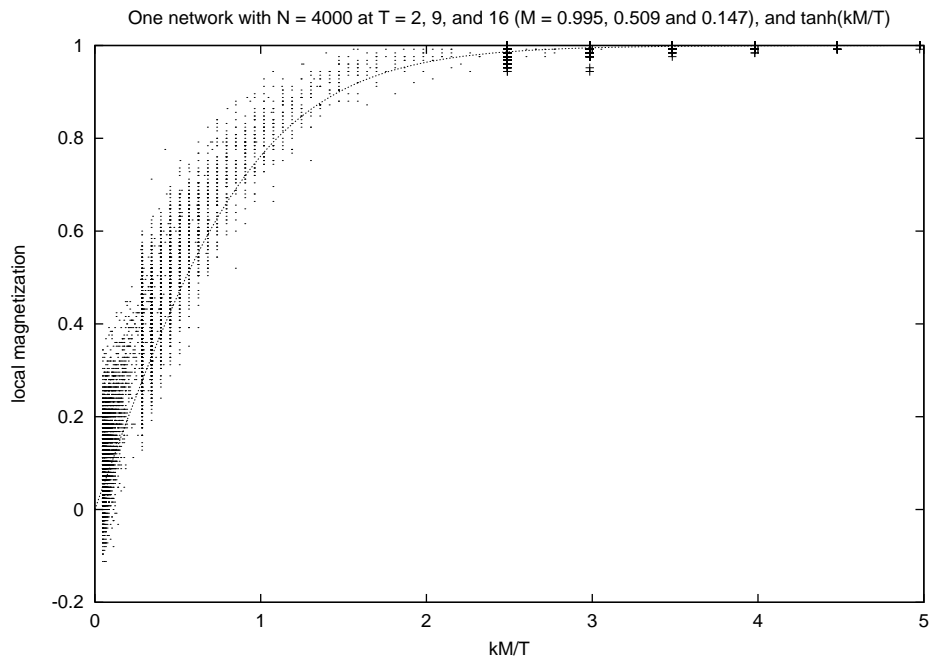}
\end{center}
\caption{
Correlation between the number of neighbours and the local magnetization $<S_i>$
for one network of $N=4000$ at $T=2,$ 9 and 16. Average over $250 < t \le 
500$ iterations. The curve is the mean field prediction tanh($\beta kM$).
$N = 2000000$ gives similar effects.
}
\end{figure}  

\begin{figure}[hbt]
\begin{center}
\includegraphics[angle=-90,scale=0.5]{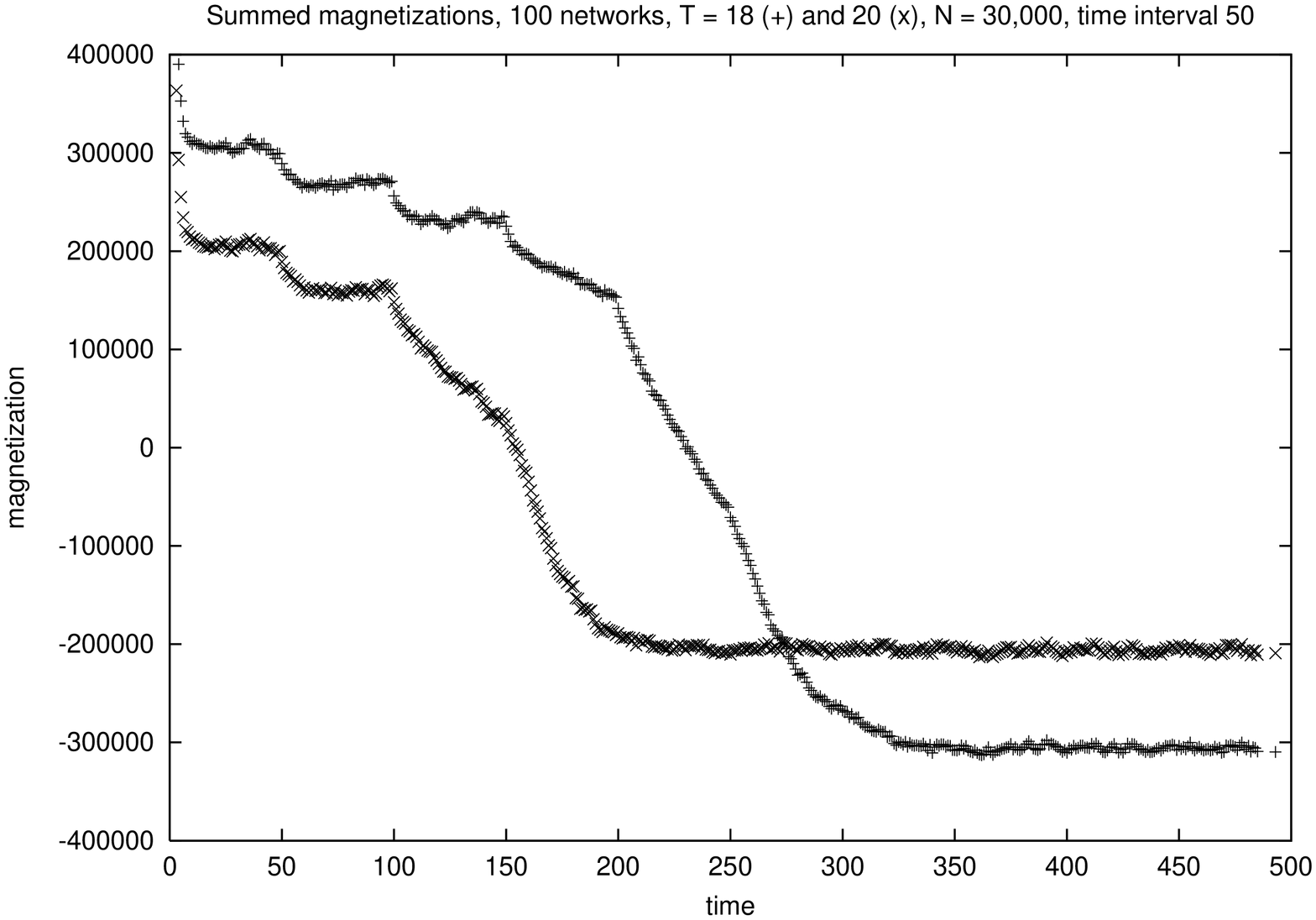}
\end{center}
\caption{
Total magnetization versus time, summed over 100 networks of $N = 30,000$ 
when after every 50 iterations the most-connected free spin is forced down 
permanently. For lower temperatures the sign change of the magnetization 
happens later. $N = 2000000$ gives similar effects.
}
\end{figure}

Thus we create a Barab\'asi-Albert network of $N$ sites added to an initial core
of $m$ fully connected sites. Each of the $N$ new nodes 
is connected to $m$ randomly selected previous nodes. 
(We allow more than one connection to the same 
site for the later added nodes; mostly we take $m=5$.) 
Then we freeze this network structure, put an 
Ising spin
$S_i = \pm 1$ onto every site, with all spins up initially. Then with the 
standard heat bath Monte Carlo algorithm spins search for thermal equilibrium
at temperature $T$ (all temperatures are given in units of coupling constant 
over Boltzmann constant). 
Fig.1a shows the resulting magnetization (averaged over the last half of 500
sweeps through the network) versus temperature; it seems to decrease 
exponentially with increasing temperature, until due to finite-size effects 
it oscillates about zero. This effective Curie temperature $T_c (N)$ is fitted
in Fig.1b onto 2.6 ln$(N) - 3$ for $5 \le N \le 5,000,000$.

Figure 2 shows that in the ferromagnetic region the spins with few neigbours
flip up and down while those with many neighbours point up most of the time.

Analogous to the appearance and spreading of new opinions in society \cite{hol}
we now try to flip the spontaneous magnetization for $T < T_c$ by
forcing the most-connected spin permanently down; then we pin in the same way
the second-most-connected spin, and so on, all in time intervals of 50
iterations. We see that in general removal of the few leading spins having
hundreds of neighbours is sufficient to flip the magnetization. (Since we
do not apply any magnetic field and have no fixed boundary conditions, we 
expect the flipping of the magnetization to be a nucleation event, which would
happen even if we flip only one randomly selected spin provided we wait long
enough.) Fig.3 shows the magnetization versus time averaged over 100 samples. 
Higher temperatures require fewer leading spins to be pinned.

Now we present a simple mean field theory for some of these effects.
Let us consider the BA network with the characteristic constant $m$ and the
corresponding Ising model with the ferromagnetic coupling constant $J=1$. The
probability that a node has a degree $k$ is given by $ P(k) \simeq Ak^{-3}$ 
when $k \gg m$  and $P(k)=0$ when $k<m$. For large networks $(N\rightarrow 
\infty)$ the normalization constant equals $A \simeq 2m^2$. In the mean
field approximation (MFA) we can simplify interactions among each group of spins
with a fixed value of $k$ by the effective field $kM$ where $M$ is the
mean magnetization (per one spin). It follows that
$$M=\int P(k)\tanh(\beta Mk) dk$$
($\beta=1/T$). This is a transcendental algebraic equation for $M(T)$. 
It is easy to find analytically the critical temperature $T_c$ that
corresponds to the case when lhs and rhs are tangent at $M=0$ .
Differentiating both sides over $M$  and putting $M=0$ we get 
$1=A \int_{m}^{\infty}\beta k^{-2}dk$;  thus $T_c=2m.$ The result can be
written as $T_c=<k> \simeq 2m = $ the mean degree of the node. This is
the typical MFA result. Fig.1 shows it to be correct in order of magnitude and
in its increase with increasing $m$. But the MFA does not describe the         
logarithmic size effect on $T_c(N)$, due perhaps to exponentially rare and
small regions of densely connected spins \cite{aleks}.

Knowing the mean magnetization $M(T)$ one can calculate the correlation $\mu(k)$
of local magnetization to the spin degree $k$. 
In fact in MFA we can write $ \mu(k)=\tanh[\beta k
M(\beta )]. $  It follows that for all temperatures $T<T_c$ this dependence is
a universal function of $u=\beta k M(\beta)$. This prediction is consistent
with figure 2. 

The effect of pinning of the most important spins (Fig. 3) can be also described
analytically and it follows that there occurs a {\bf discontinuous} phase
transition from the "spin up" phase to the "spin down"  phase by a well defined
critical number of pinned spins.

Pinning one spin of degree $k$ to the state $S=-1$ is equivalent in MFA 
to lowering the mean magnetization by $\mu(k)/N$  and introducing to the system 
the {\it external} magnetic field of the magnitude $J=1$
that is oriented "down". This field is felt {\bf only} by $k$ other spins thus
its mean value for the whole system equals $b(k)=-k/N$. If we pin $j$ of the
most connected spins then it means that we pin all spins of the degree 
$k > \kappa$ where $$ j=NA\int_{\kappa}^{\infty}k^{-3}d k\quad . $$ 
It follows that  we decrease the mean
magnetization (per one spin)  by $$\delta M(j)=
A\int_{\kappa}^{\infty}(\mu(k)+1)k^{-3}d k\simeq \frac{2j}{N}  $$ where we
have assumed that all pinned spins were completely ordered before pinning,
$\mu(k)=1$. The  effective {\it internal} field $B(j)$ acting in the system from
the pinned spins can be calculated as $$ B(j)=- A\int_{\kappa}^{\infty}
\mu(k)k^{-2}d k \simeq -2m\sqrt{j/N} \quad .$$ It is important to stress that 
$\delta M \propto 1/N$ but $B(j) \propto \sqrt{1/N}$ and this internal
field causes much larger decrease of magnetization $\Delta M$ than the direct
effect of pinning. For small $j$ we have  $\Delta M(j) =\chi B(j)$ where $\chi$
is the system initial susceptibility. In MFA it can be calculated as the mean
value of the derivative of local magnetization $\mu(k)$ over the field $B(j)$
and it equals to $$ \chi= \frac{2m^2}{T} \int_{m}^{\infty} \frac{d k}{k^3
\cosh^2(\beta Mk)}\quad .$$ The last integral cannot be calculated analytically.
Pinning  several spins means  influencing the system by a large internal field
$B(j)$ that can even cause a flip of the total magnetization. The value for
the minimal flipping field $B^*$ can be obtained from the equation $$
M^*=A\int_m^\infty k^{-3}\tanh[\beta(M^* k +B^*)] d k $$ and the corresponding
tangency relation at the critical point $M^*$: $$ 1=\beta A\int_m^\infty
k^{-2}\cosh^{-2}[\beta(M^*k+B^*)]dk \quad. $$The above equations form imply
conditions for $B^*$ and $M^*$. After some algebra one can find the following
relation between $B^*$ and $M^*$: $$B^*=-mM^*+T\,{\rm artanh}(M^*/2).$$Combining
the last result with the relation for $B(j)$ we get the following equation for
the minimal number $j^*$ of pinned spins needed to invert the system 
magnetization:
$$\sqrt{\frac{j^*}{N}}= \frac{M^*}{2}-\frac{T}{2m}{\rm artanh}\frac{M^*}{2} 
\simeq (M^*/2)(1-T/T_c)$$
where the last approximation is valid only for the small $M^* = 0.03 \dots 0.1$
of Fig.3 and we used the above $T_c=2m$. This prediction 
$j^*/N \sim 10^{-4} \dots 10^{-3}$ agrees reasonably with our simulations. 

\bigskip
In summary, we combined the Ising model with the Barab\'asi-Albert network and
found that depending on the convincing power one either has a majority opinion
or two equally widespread opposing opinions, i.e. a Curie point. 

We thank DAAD and the special grant of Warsaw University of Technology
to support the visit to Cologne
University, the J\"ulich supercomputer center for time on their Cray-T3E,
and A.L. Barab\'asi for information about similar work.

\end{document}